\documentclass[aps,english,prl,twocolumn,showpacs]{revtex4-1}

\usepackage{graphicx}
\usepackage{amssymb}
\usepackage{babel}
\usepackage{epstopdf}
\usepackage{hyperref}

\begin{document}

\title{Electric breakdown effect in the current-voltage characteristics of amorphous indium oxide thin films near the superconductor-insulator transition}

\author{O.~Cohen, M.~Ovadia, and D.~Shahar}
\affiliation{Department of Condensed Matter Physics, Weizmann Institute of Science, Rehovot 76100, Israel}

\date{September 16, 2011}

\begin{abstract}
Current-voltage characteristics in the insulator bordering superconductivity in disordered thin films exhibit current jumps of several orders of magnitude due to the development of a thermally bistable electronic state at very low temperatures. In this high-resolution study we find that the jumps can be composed of many (up to 100) smaller jumps that appear to be random. This indicates that inhomogeneity develops near the transition to the insulator and that the current breakdown proceed via percolative paths spanning from one electrode to the other.

\end{abstract}


\maketitle

Disordered, thin-film superconductors can be driven into an insulating state by varying an experimental parameter such as the level of disorder or the magnetic field ($B$) (for a review, see Ref. \citep{gantmakher2010superconductor}). The $B$-driven insulating state is nonmonotonic, the resistance showing a pronounced peak as a function of $B$ \citep{PhysRevLett.69.1604,PhysRevB.69.024505,PhysRevLett.92.107005,PhysRevLett.98.127003}. At $B$'s much beyond the peak the insulating behavior is replaced by a weakly insulating, or perhaps metallic \citep{gantmakher1998destruction,PhysRevLett.98.127003}, state. Several theoretical groups have addressed this nonmonotonic insulator \citep{PhysRevLett.65.923,PhysRevLett.95.077002,PhysRevB.73.054509,2008Natur.452..613V,PhysRevB.79.134504,PhysRevLett.105.267001}, but a consensual framework is yet to emerge.

One key observation in both the disorder-driven and $B$-driven insulators is that the current-voltage ($I$-$V$) characteristics exhibit a voltage threshold $V_{th}$, separating high-resistance (HR) and low-resistance (LR) states. At low temperatures ($T<0.2$ K), $V_{th}$ is accompanied by a large (several orders of magnitude) current jump, as well as hysteresis \citep{PhysRevLett.94.017003,PhysRevLett.99.257003}. This switching behavior was interpreted in light of a model \citep{PhysRevLett.102.176802,PhysRevLett.102.176803,2010arXiv1004.5153K} in which Joule heating by the current flow can induce a nonequilibrium state where the electrons' $T$, $T_{el}$, can significantly exceed the phonon bath $T$. One consequence of this model is that, at very low $T$, the electrons enter a bistable region where, as a function of voltage, $T_{el}$ can abruptly change causing the large current jumps between the ``cold" HR state and the ``hot" LR one. However, there have been some indications to suggest that this is not the whole picture. In recent work on insulating TiN thin films, multiple conductance steps in the $dI/dV$ curves were observed near the threshold at low $B$'s \citep{Baturina2008316}. This preliminary finding was taken as an indicative of a percolative behavior, reflected by multiple charge depinning transitions \citep{PhysRevLett.100.086805}, with the threshold behavior later attributed to heating instability \citep{2010arXiv1004.5153K}.

In this Rapid Communication we report the results of high-resolution $I$-$V$ measurements in the vicinity of $V_{th}$. These more detailed studies reveal the existence of several (as many as 100) discontinuous $I$ steps, accompanied by hysteresis, near $V_{th}$. A thorough experimental analysis of these multisteps supports a scenario in which the current distribution develops significant inhomogeneity near the superconductor-insulator transition (SIT): Each $I$ step corresponds to a sudden appearance or disappearance of a narrow conductive path of ``hot" electrons that carries relatively high current through the ``cold" insulator.

The system we study consists of highly disordered amorphous indium oxide ($a$:InO) that was fabricated by $e$-gun evaporation of $99.99\%$ pure sintered In$_{2}$O$_{3}$ pieces onto oxidized silicon wafers \citep{0022-3719-19-26-018,PhysRevB.46.10917}. Premade Au leads and a lift-off mask were used to pattern devices with a well-defined two-terminal geometry. $25$$-$$30$ nm thick and $500$$-$$1000$ $\mu$m long square films were used throughout this study. In addition, one hall-bar sample deposited through a shadow mask and contacted by hand-pressed indium was used to test the influence of the contact resistances on the results. Some of the samples were heat treated at $40$ $^{\circ}$C in vacuum for a few hours postdeposition to reduce their degree of disorder by annealing \citep{PhysRevB.46.10917}. All measurements were performed using a standard two-probe technique inside a dilution refrigerator capable of reaching $0.02$ K. The $I$-$V$ curves were traced by sweeping dc voltage and monitoring the current with a low-noise transimpedance amplifier. The resistance of the devices was measured by a low-frequency ($0.3$$-$$3$ Hz) lock-in configuration, or by the low-voltage derivative of the $I$-$V$ characteristics.

\begin{figure}

\includegraphics[width=1\columnwidth]{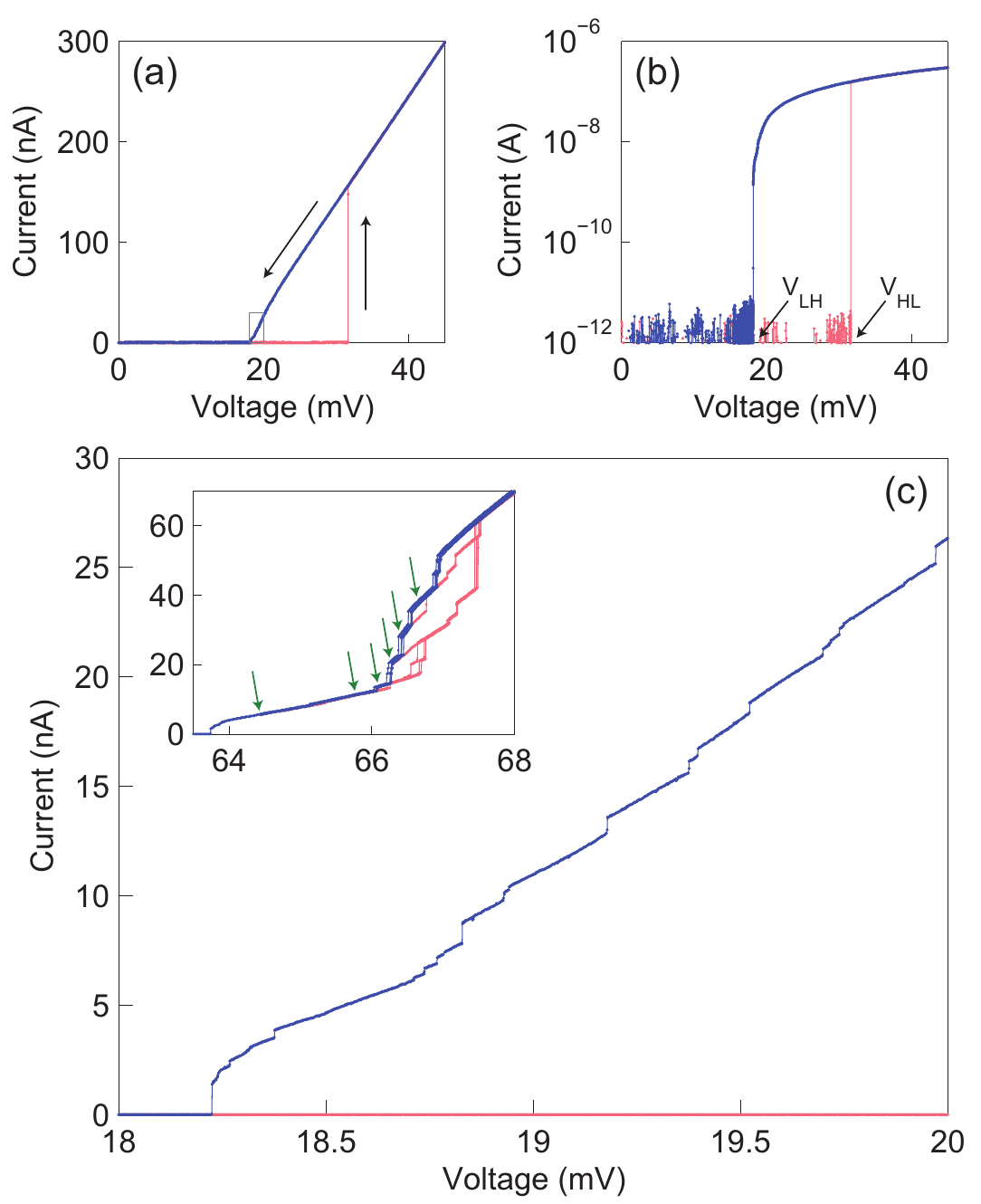}

\caption{(a) $I$-$V$ characteristic of sample LW1a at $T=70$ mK and $B=2$ T. The arrows show the voltage sweep direction; the positive sweep is shown in red, while the negative sweep is in blue. This color scheme will be followed in the subsequent figures. (b) Same as (a) plotted on a semilogarithmic scale. The threshold voltages $V_{HL}$ and $V_{LH}$ are marked by the two arrows. (c) Closeup view on the gray rectangle in (a). Inset: $I$-$V$ characteristics of LW3f at $T=30$ mK and $B=9$ T. Each green arrow denotes a turning point of the sweep direction.}

\label{introIV}
\end{figure}

A typical switching behavior in our samples can be seen in Fig. \ref{introIV}(a), where we present a $T=70$ mK $I$-$V$ curve taken from sample LW1a. At low $V$, $I$ through the film is small ($<1$ pA), inapparent on the scale of the plot. As $V$ reaches a certain voltage, referred to as $V_{HL}$, conduction starts with $I$ increasing abruptly by several orders of magnitude. On reversing the sweep direction, hysteresis is observed and the LR state persists until a second voltage, $V_{LH}$, is reached. The switching behavior is even more pronounced when plotted using a logarithmic ordinate [Fig. \ref{introIV}(b)], better illustrating the range of variation in $I$ at the two thresholds.

The central result of this work is shown in Fig. \ref{introIV}(c). Focusing on the data obtained from the negative sweep direction close to $V_{LH}$ reveals a series of discontinuous steps. These steps seem random in location, as well as jump magnitude, but are highly reproducible after successive scans, even when the two measuring contacts are interchanged. The step at the lowest $V$ corresponds to the threshold we designated as $V_{LH}$, beyond which the system reverts to the HR state.

The $I$ steps are accompanied by a complex hysteresis phenomenon. This is manifested in the inset of Fig. \ref{introIV}(c), which shows a set of $I$-$V$ curves obtained by cycling $V$ between a voltage higher than the threshold $V_{HL}$ and a voltage just below each step. Several distinct hysteresis loops are seen in these data, each associated with a specific current step. While this behavior is observed under all experimental parameters for the $I$ steps to appear, the exact dependence of the subhysteresis on these parameters remain to be investigated further.

\begin{figure}

\includegraphics[width=1\columnwidth]{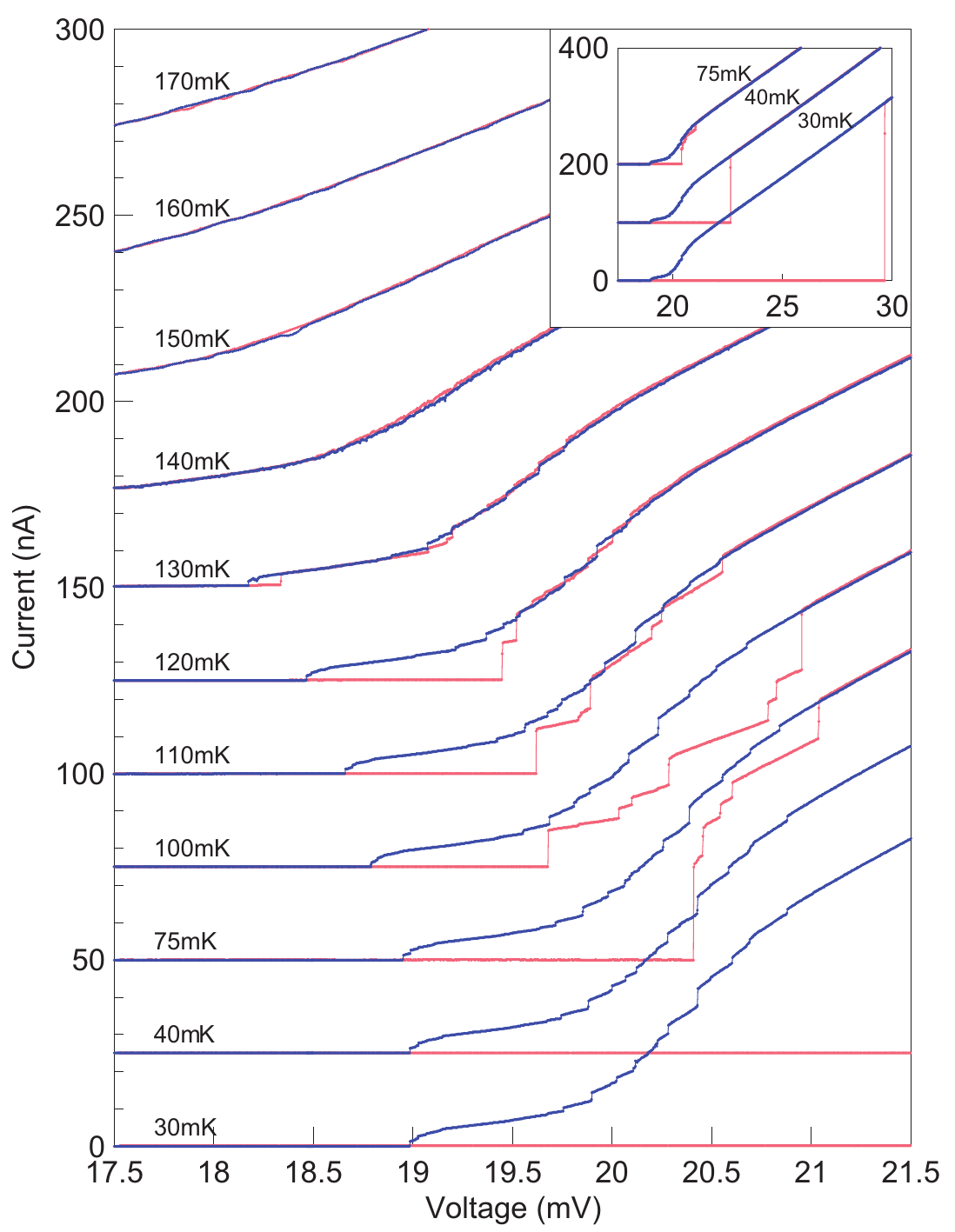}

\caption{$I$-$V$ isotherms of LW3f at the vicinity of the threshold, taken at $B=4$ T. For clarity, the curves are offset from one another by $25$ nA. Inset: Expanded $V$ range view of the $30$, $40$, and $75$ mK scans, showing the shifting of $V_{HL}$.}

\label{tempIV}

\end{figure}

Figure \ref{tempIV} shows the $T$ evolution of the $I$ steps measured across film LW3f. As $T$ increases, the steps pattern in the negative sweep branch deforms gradually, the location of each step shifts to lower voltages, and the step magnitude decreases until the discontinuity vanishes altogether.

Interestingly, multiple $I$ steps appear also in the positive sweep direction (red traces in Fig. \ref{tempIV}). While at very low $T$ we observe a single step at $V_{HL}$ (see the inset of Fig. 2 for a wide $V$ range), $V_{HL}$ rapidly decreases with $T$ and,  above $\sim50$ mK, breaks into multiple steps similar to those in the negative sweep direction. Eventually, the hysteresis collapses and the two branches merge into one.

\begin{figure}

\includegraphics[width=1\columnwidth]{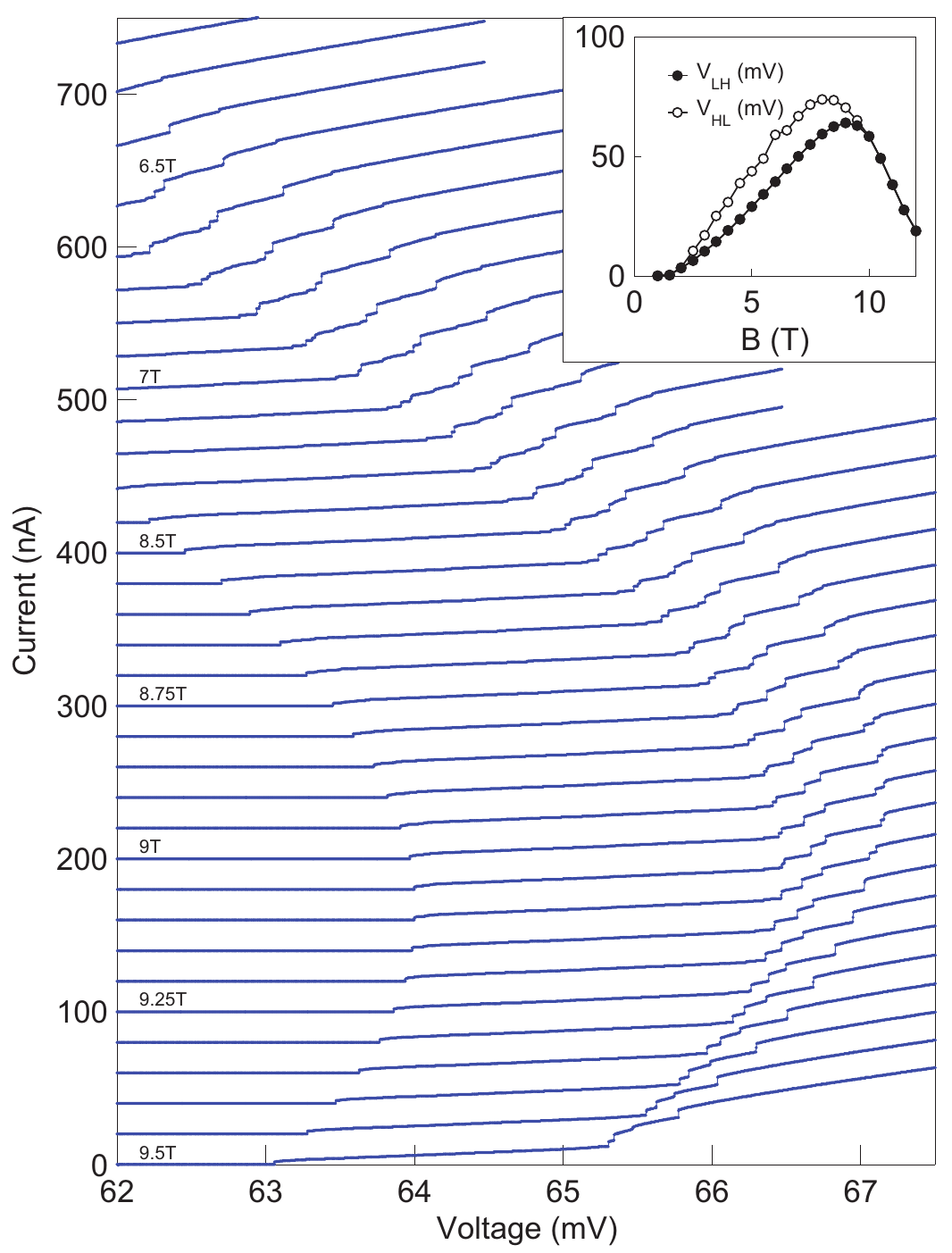}

\caption{(a) $I$-$V$ curves of LW3f as a function of $B$ at the vicinity of the threshold, taken at $T=30$ mK. Only negative sweep data are shown. For clarity, an offset current of multiples of $20$ nA has been added. Inset: $V_{LH}$ and $V_{HL}$ vs $B$ of the same sample.}

\label{magneticIV}

\end{figure}

The evolution of the steps pattern as function of $B$ is shown in Fig. \ref{magneticIV}, where we plot, for sample LW3f, the dependence of the $I$-$V$ characteristics on perpendicular $B$. The critical point of the $B$-driven SIT for this sample is $B_{C}=0.9$ T and the magnetoresistance (MR) maxima was found to occur at $5$ T field. The step profile appears only for fields in the range $B_{C}<B\lesssim10$ T, beyond which the curve is smooth. The location of each step moves to higher $V$'s as $B$ is increased, and reaches a peak near $B=9$ T. Simultaneously, as the steps shift, their height increases, reaching a maximum value, and then decreases to zero. The corresponding $B$ dependence of $V_{HL}$ and $V_{LH}$ is given in the inset. As the difference $V_{HL}-V_{LH}$ approaches zero, near $B_{C}$ and near $10$ T, the hysteresis closes and multiple steps are observed also in the positive sweep direction (data not shown).

The magnetic field is seen to have only a moderate influence on the hierarchy of the step pattern. For each $I$ step, a trajectory on the $B$-$V$ plane can be clearly and independently traced from just above $B_{C}$ to its eventual high-$B$ demise. Even though these trajectories differ in shape, they seldom appeared to cross one another, leaving the step arrangement almost unchanged over a wide range of $B$'s.

\begin{figure}

\includegraphics[width=1\columnwidth]{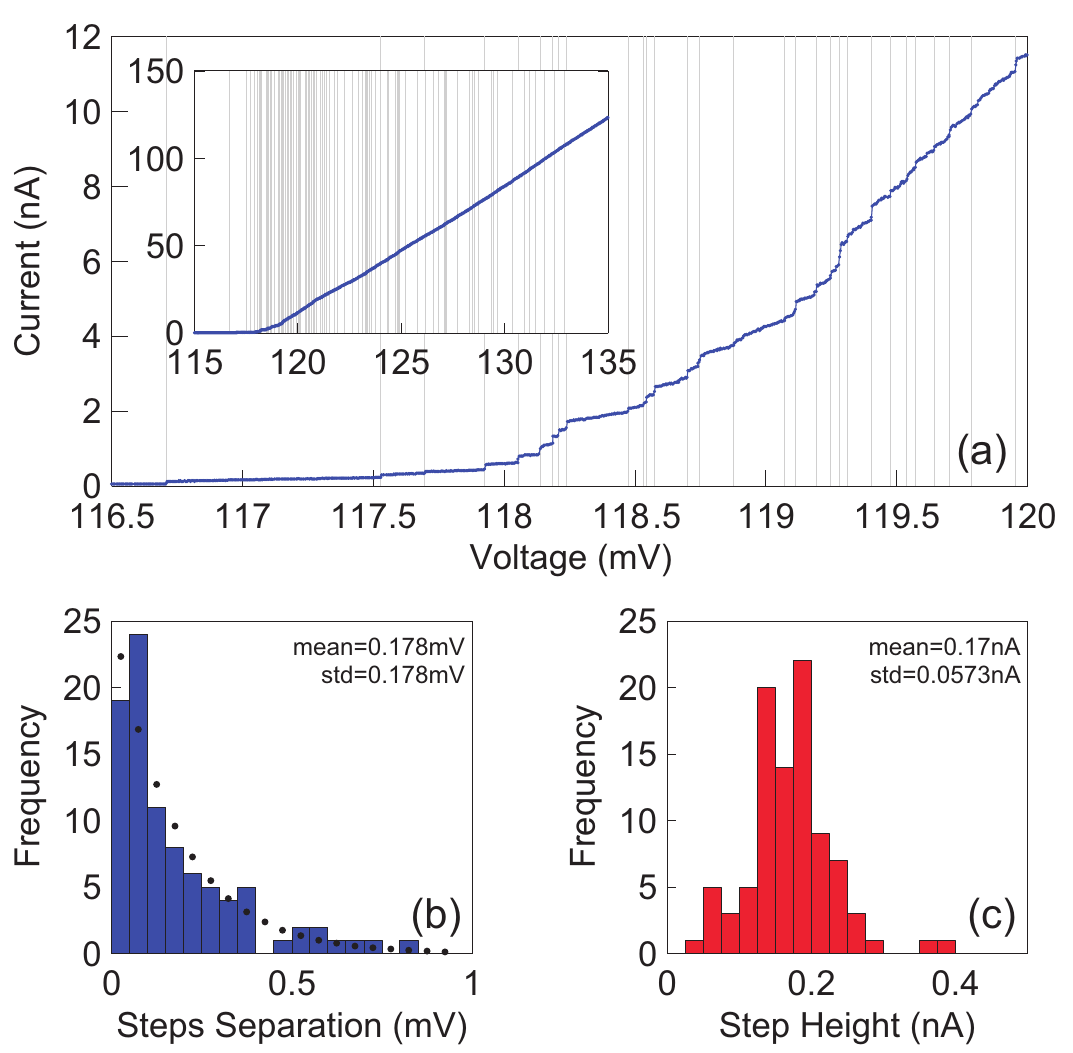}

\caption{$I$-$V$ curve of LW3d at $T=20$ mK and at $B=0$. The step locations are marked by the gray vertical lines. Inset: Wide $V$ range view of the same curve. (b) Histogram showing the distribution of the $V$ differences between consecutive steps. The black dots are of the exponential distribution of the same mean. (c) Histogram of the step height distribution.}

\label{disorderIV}

\end{figure}

We now turn to films with a higher degree of disorder. Sample LW3b has a disorder value that is larger than critical and is insulating at $B=0$. Nonetheless, it still exhibits a MR peak at $B=2$ T \citep{1996JETP...82..951G,PhysRevLett.99.257003,PhysRevB.78.014506,PhysRevLett.106.127003}. In Fig. \ref{disorderIV}(a) the $I$-$V$ step pattern of this device at $T=20$ mK and $B=0$ is shown. A few dozens $I$ steps can be seen within the sensitivity of the measurement, almost an order of magnitude more than for the superconductivity samples. The step locations, found using an edge detection algorithm, are delineated by the gray vertical lines. In total, $92$ $I$ steps were detected.

The large number of steps enables us to carry a statistical analysis of the process. Using the data in Fig. \ref{disorderIV}(a), we plot two histograms showing the $V$ differences between consecutive steps and the step heights. The $V$ difference histogram is exponentially distributed, suggesting that the steps are randomly spread in a Poisson-like fashion. The step-height histogram is bell shaped, centered on a mean height of $0.17$ nA with a low standard deviation value of $0.057$ nA. For comparison, this analysis was performed on the same sample after thermal annealing for 72 h (at its postannealing state, this sample was referred to above as LW3f). The corresponding total number of steps and their mean height, extracted from an $I$-$V$ curve at $T=30$ mK and $B=4$ T (Fig. \ref{tempIV}), is $22$ and $1.21$ nA, respectively.

To account for our findings we adopt the perspective that transport in these films can become spatially inhomogeneous. In this context our results appear quite reasonable, considering that the conduction transition is not singular and involves multiple events of the same sort, randomly arranged, and strongly contingent on the level of disorder. Further evidence for such inhomogeneity near the SIT have been reported in the past \citep{Kowal1994783,PhysRevB.59.11209,PhysRevLett.92.107005,PhysRevB.74.100507,PhysRevLett.99.257003,PhysRevLett.101.157006,sacepe2011localization}, and has even provided a basis for several models describing the phase transition \citep{PhysRevB.73.054509,PhysRevLett.100.086805,2008Natur.452..613V,PhysRevB.79.134504,PhysRevLett.81.3940,dubi2007nature}. If the fluctuations in conductivity are strong, the current flows along narrow, percolating, channels of least resistance. Combining this view with the electron heating scenario from Refs. \citep{PhysRevLett.102.176802,PhysRevLett.102.176803} gives rise to the possibility of local overheating of electrons at each path.

Within this physical picture, we can attribute the $I$ steps to spatially separated bistabilities in $T_{el}$. Depending on the applied $V$, the conduction regime of each channel can be toggled almost individually between the ``cold" HR state and the ``hot" LR one. The low-$T$ behavior can be understood as follows: At low $V$, all channels are nearly closed. When $V_{HL}$ is reached, conduction begins with the opening of all channels in an avalanche fashion; the first channel to open triggers the conduction in nearby channels, which then proliferates to the bulk of the sample. In the $I$-$V$ curve, this process appears as one big current step. Upon reducing $V$, channels close one by one, and so only small steps are found in the opposite sweep direction. This behavior varies somewhat at high $T$, where, as depicted in Fig. \ref{tempIV}, we find that small steps also appear in a positive sweep direction, indicating that the high-$T$ rupture of the ``hot" phase is also successional.

From this interpretation, it follows that the $I$ step pattern serves as a fingerprint of the particular configuration of inhomogeneities. Varying the degree of disorder considerably influences the low-$T$ inhomogeneity: At lower disorder fewer channels are available, but a higher current flows through each one. On the other hand, the hierarchy of the steps, being largely conserved over a wide range of $B$'s, suggests that inhomogeneity is only weakly dependent on $B$. This is intriguing, considering the strong field dependence of the threshold, and certainly calls for a more thorough theoretical analysis.

It is interesting to note the similarities between insulating $a$:InO and polycrystalline TiN bordering the SIT. Shared key features include the Arrhenius-activated behavior, the pronounced peak in the MR \citep{PhysRevLett.92.107005,PhysRevB.69.024505}, and the high-field saturation \citep{gantmakher1998destruction,PhysRevLett.98.127003}. When comparing the present data with those from Ref. \citep{Baturina2008316}, additional distinct similarities are observed. Both materials exhibit a set of sharp steplike $V$ thresholds that are nonmonotonic in $B$ and attain their maximal value far above the MR peak, further suggesting a common underlying mechanism.

In summary, the appearance of multiple $I$ steps near the threshold for conduction can be linked to an inhomogeneous conduction state that develops in the vicinity of the SIT. We note that features similar in appearance to those reported here have been observed in the context of two-dimensional quantum dot arrays \citep{PhysRevLett.74.3237}, a system in which current is known to be inhomogeneously distributed due to disorder \citep{PhysRevLett.71.3198}. These similarities further advocate the relevance of Coulomb blockade in the SIT insulating phase \citep{Kowal1994783,PhysRevB.73.054509}, and prompt the use of quantum dot arrays as a model system for understanding its transport properties.

We wish to thank V. Kravtsov, Y. Meir, and B. Sac\'ep\'e for illuminating discussions. This work was supported by the Minerva foundation with funding from the Federal German Ministry for Education and Research.


\end{document}